\documentclass[]{mn2e}
\usepackage{graphicx}
\usepackage{amsmath}

\title[Scale Transformation and the Bispectrum]
{Scale Transformations, Tree-level Perturbation Theory, and the
Cosmological Matter Bispectrum}
\author[Pan, Coles \& Szapudi]{Jun Pan$^{1, 2}$, Peter Coles$^{2,3}$ and Istv\'an Szapudi$^4$
\\$^1$Purple Mountain Observatory, Chinese Academy of Sciences, Nanjing 210008, China
\\$^2$School of Physics and Astronomy, University of Nottingham, Nottingham NG7 2RD, UK
\\$^3$School of Physics and Astronomy, Cardiff University, Queeens
Buildings, The Parade, Cardiff CF24 3AA, UK
\\$^4$Institute for Astronomy, University of Hawaii, Honolulu HI 96822, USA}

\begin{document}

\maketitle


\begin{abstract}
Scale transformations have played an extremely successful role in
studies of cosmological large-scale structure by relating the
non-linear spectrum of cosmological density fluctuations to the
linear primordial power at longer wavelengths. Here we generalize
this approach to investigate the usefulness of scale transformations
for nonlinear higher-order statistics, specifically the bispectrum.
We find that the bispectrum predicted by perturbation theory at
tree-level can be rescaled to match the results of full numerical
simulations in the weakly and intermediately nonlinear regimes,
especially at high redshifts, with an accuracy that is surprising
given the simplicity of the procedure used. This discovery not only
offers a simple practical way of calculating the matter bispectrum,
but also suggests that scale transformations may yet yield even
deeper insights into the physics of hierarchical clustering.
\end{abstract}


\section{Introduction}

In the linear stage of gravitational instability, cosmological
density perturbations evolve in such a way that each Fourier mode
grows at the same rate as the others, preserving the shape of the
primordial power spectrum. In the later stages, things become much
more complicated, with mode-mode interactions changing the relative
growth rates of different harmonics and coupling their phases
together to generate non-zero higher-order spectra (Coles \& Chiang
2000). Nevertheless, inspired by the creative work of Hamilton et
al. (1991), empirical formulae have been established that can
accurately predict the two-point correlation function $\xi(r)$ or
power spectrum $P(k)$ of cosmological density fluctuations in the
non-linear regime given only the initial linear power spectrum and
the background cosmological model (Jain, Mo \& White 1995; Peacock
\& Dodds 1996). The key element of these formulae is a scale
transformation which expresses the conservation of particle pairs
which, in turn, is one of the infinite set of equations that forms
the BBGKY hierarchy, c.f. Peebles (1980). The discovery of this
unexpectedly successful transformation for the nonlinear $\xi(r)$ or
$P(k)$ was a great leap forward in the understanding of cosmological
clustering evolution, but with the renaissance of halo model (Cooray
\& Sheth 2002), and the emergence of various high-level perturbative
techniques (e.g. Szapudi \& Kaiser 2003;  McDonald 2007), interest
in the classic form introduced by Hamilton et al. (1991) has to some
extent faded away. Nowadays it seems much more fashionable to
calibrate gravitational nonlinearity by tuning parameters of the
halo model or using empirical fitting formulas based on it (Smith et
al. 2003).

After great successes with two-point correlation functions, it was
widely anticipated that the halo model would be able to generate
precise nonlinear three-point statistics over a full range of length
scales (Scoccimarro et al. 2001; Takada \& Jain 2003). Detailed
examination against simulations, however, has proved somewhat
disappointing. New elements have had to be incorporated to improve
the current halo model even after introducing {\em ad hoc} free
parameters accounting for mass cutoff and halo boundary (Fosalba,
Pan \& Szapudi 2005), or replacing spherical haloes with triaxial
ones (Smith, Watts \& Sheth 2006).

In this paper, we revive the scale transformation idea to derive a
model that approximates the nonlinear contributions to the
bispectrum of cosmological matter perturbations. The bispectrum is
the lowest order (and therefore the simplest) diagnostic of
non-Gaussianity because it vanishes identically for any Gaussian
random field regardless of its power spectrum. If the initial
density fluctuations are Gaussian, as is expected in the simplest
inflationary models, the bispectrum is therefore completely induced
by non-linear gravitational evolution. However, these higher-order
effects on the bispectrum are difficult to calculate with reasonable
accuracy even on large scales where the evolution is quasi-linear.
Differences between full numerical simulations and the predictions
of  Eulerian perturbation theory at tree-level exceed $\sim 10-20\%$
at a scale $k \sim 0.1h/$ Mpc. The perturbation theory at one-loop
level, the next possible improvement on the tree-level theory,
requires heavy numerical integration and has limited success
(Scoccimarro et al. 1998). Lagrangian perturbation theory at second-
and third-order provides a better template than Eulerian theory, but
its dynamic range is heavily restricted by the shell--crossing scale
at around $k\sim 0.4h/$ Mpc (Scoccimarro 2000). A further
inconvenience of adopting Lagrangian perturbation theory is that the
bispectrum it predicts is not {\em computed}, but {\em measured}
from N-body simulations of which the particle motions are controlled
by Lagrangian theory.

One practical way forward is to modify the kernel of perturbation
theory at tree-level to develop empirical fitting formulae such as
 extended and the hyper-extended perturbation theory (Colombi et al.
 1997; Scoccimarro \& Frieman 1999). However, the performance of
 these techniques known to be quite poor, even in cases where the initial power-spectrum is completely
scale-free (Hou et al. 2005). The best results so far in this vein
relate to the empirical model of Scoccimarro \& Couchman (2001,
herefter SC2001), using hyper-extended perturbation theory, but its
average deviation from $N$-body results is at the level of $15\%$,
still way beyond the accuracy levels expected in the era of
precision cosmology. On the other hand, even this produces some
dubious features, such as the apparently trough in the bispectrum at
scales $\sim 0.03<k<\sim 0.12h/$ Mpc, where baryon acoustic
oscillations leave their footprint on large-scale structure
statistics.

An accurate template for the bispectrum on large scales is extremely
important in the epoch where large galaxy redshift surveys are
available to unveil details of large scale structure and to
constraint cosmological parameters with higher-order statistics. It
is possible that an accurate model can be achieved based on the
formula of SC2001 facilitated with high precision simulations in
huge box, or perhaps with a theory  established by the
renormalization technique of McDonald (2007). However, in this paper
we will show that, at least in the quasi- and intermediate nonlinear
regime, it is possible to recover the bispectrum quite accurately
using the scale transformation described above, an approach which is
free from the laborious calibration of fitting parameters and
formidable multi-dimensional integrations.

\section{Scale Transformations and the Bispectrum}

\subsection{Scale Transformations of Second-order Statistics}
Let $\xi(r)$ be the two point correlation function in real space at
scale $r$. The number of neighbours of a central point within a
spherical top-hat window of radius $r$ is given by
$r^3[1+\overline{\xi}(r)]$, where $\overline{\xi}(r)=3\int_0^r
s^2\xi(s)ds/r^3$ is the volume-averaged two point correlation
function. Hamilton et al. (1991) proposed that by pair conservation
at any stage of gravitational evolution, one can quote a Lagrangian
separation $r_L$ defined by
\begin{equation}
r_L^3=r_{NL}^3\left[ 1+\overline{\xi}(r_{NL}) \right]\ ,
\end{equation}
at which the nonlinear function $\overline{\xi}_{NL}$ is an
universal function of the linear one:
$\overline{\xi}_{NL}(r_{NL})=f[\overline{\xi}_L(r_L)]$. The
effectiveness of the paradigm is at least partly explained by
Nityananda \& Padmanabhan (1994): such a nonlinear scaling relation
can arise if the pairwise velocity demonstrates certain scaling
features. In fact, based on the universal properties of the pairwise
velocity distribution, the pair conservation equation can be solved
with an iterative technique to produce the correct nonlinear $\xi$
from the input linear function (Caldwell et al. 2001).

In Fourier space, the same sort of scaling can be applied to the
dimensionless power spectrum $\Delta^2(k)=P(k)k^3/(2\pi^2)$ by a
relation of the form
\begin{equation}
k_L=\left[1+\Delta_{NL}^2(k_{NL}) \right]^{-1/3} k_{NL} \ ,
\label{eq:nsr}
\end{equation}
so that the nonlinear power spectrum $P_{NL}(k_{NL})$ is obtained
through $\Delta^2_{NL}(k_{NL})=f\left[ \Delta^2_L(k_L)\right]$. The
scale transformation of Eq.~(\ref{eq:nsr}) does not have such a firm
theoretical motivation as its real space version, although there is
no question that works very well in practice (Peacock \& Dodds
1996).

\subsection{The Bispectrum at Tree-level}
The bispectrum is the three point correlation function defined in
Fourier space. Denote the Fourier transformation of the cosmic
density contrast with $\delta({\bf k})$. The bispectrum is
\begin{equation}
B({\bf k}_1, {\bf k}_2, {\bf k}_3)=\langle \delta({\bf
k}_1)\delta({\bf k}_2)\delta^*({\bf k}_3) \delta_D({\bf k}_1+{\bf
k}_2+{\bf k}_3) \rangle \ ,
\end{equation}
in which $\delta_D$ is the Dirac function. In a statistically
isotropic universe, $B({\bf k}_1, {\bf k}_2, {\bf k}_3=-{\bf
k}_1-{\bf k}_2)$ is written as $B(k_1, k_2, k_3)$, with
$k_3=\sqrt{k_1^2+k_2^2+2 k_1 k_2 \mu_{12}}$ and $\mu_{12}={\bf k}_1
\cdot {\bf k}_2/(k_1 k_2)$.

Because the initial distribution of dark matter is Gaussian, there
is no non-zero bispectrum in the scheme of linear evolution. The
first non-trivial contribution to bispectrum, at tree-level, comes
from the second order terms in the expansion of the cosmic density
fluctuation. Explicitly,
\begin{equation}
B_{PT}(k_1, k_2, k_3)=P_L(k_1)P_L(k_2) F_2(k_1, k_2, \mu_{12}) +
{\rm cyc.} \ ,
\end{equation}
where {\rm cyc.} refers to the other two terms obtained by making
cyclic permutations of the indices of the first term, $P_L$ is the
linear power spectrum, and $F_2$ is the kernel of second order in
Eulerian perturbation theory (Goroff et al. 1986):
\begin{equation}
F_2(k_1, k_2, \mu_{12})=\frac{10}{7}+\mu_{12}\left( \frac{k_1}{k_2}+\frac{k_2}{k_1} \right)
+\frac{4}{7}\mu_{12}^2\ .
\end{equation}
Note the dependence on cosmological parameters of $F_2$ is extremely
weak, so it will be ignored here.

\begin{figure*}
\resizebox{\hsize}{!}{\includegraphics{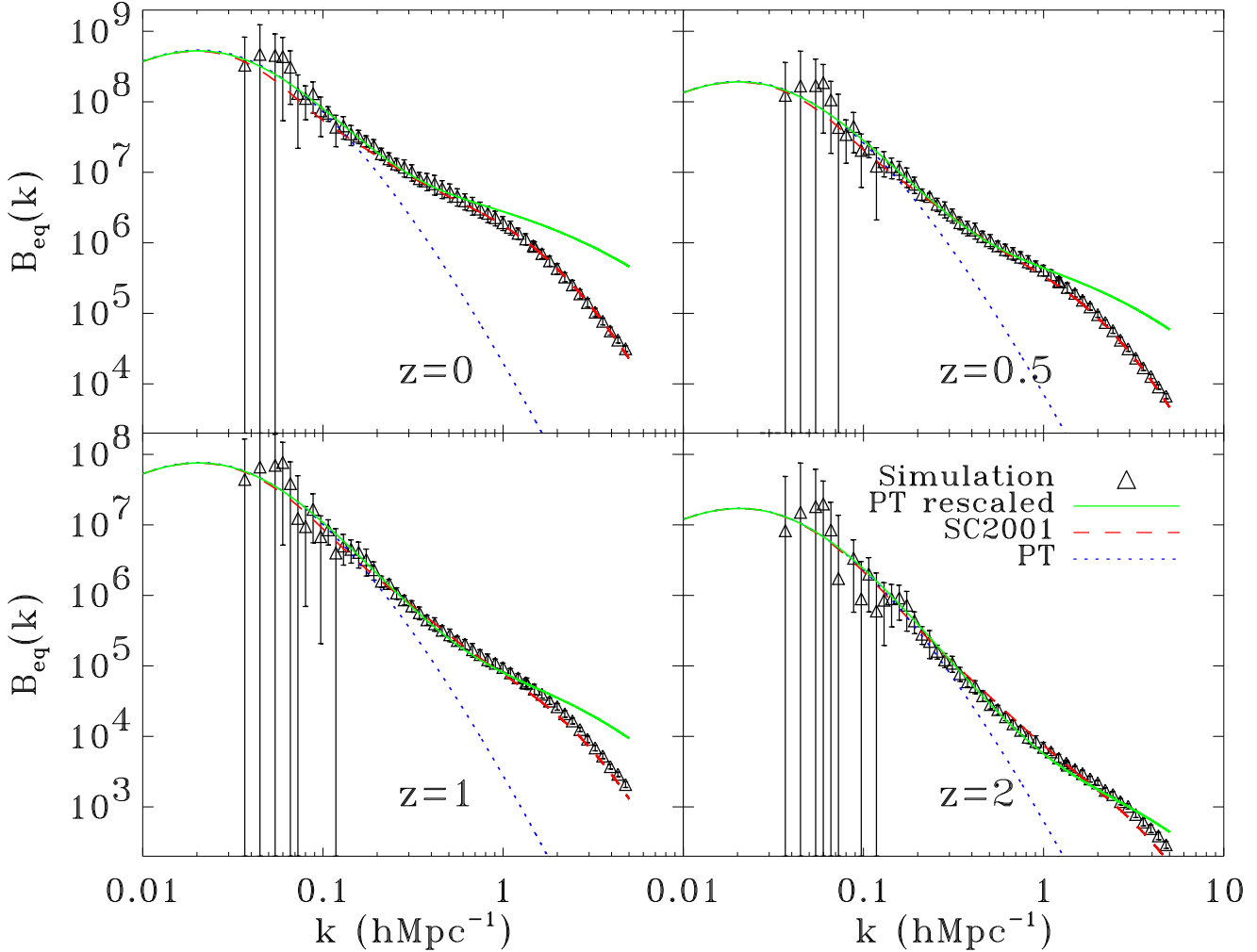}}
\caption{Bispectrum of equilateral triangles, PT means the
second-order perturbation theory, SC2001 is the prediction of
Scoccimarro \& Couchman (2001), PT rescaled is the expectation given
by Eq. (10). The spurious trough at large scales of SC2001 can be
seen in the z=0 panel.}
\end{figure*}

\begin{figure*}
\resizebox{\hsize}{!}{\includegraphics{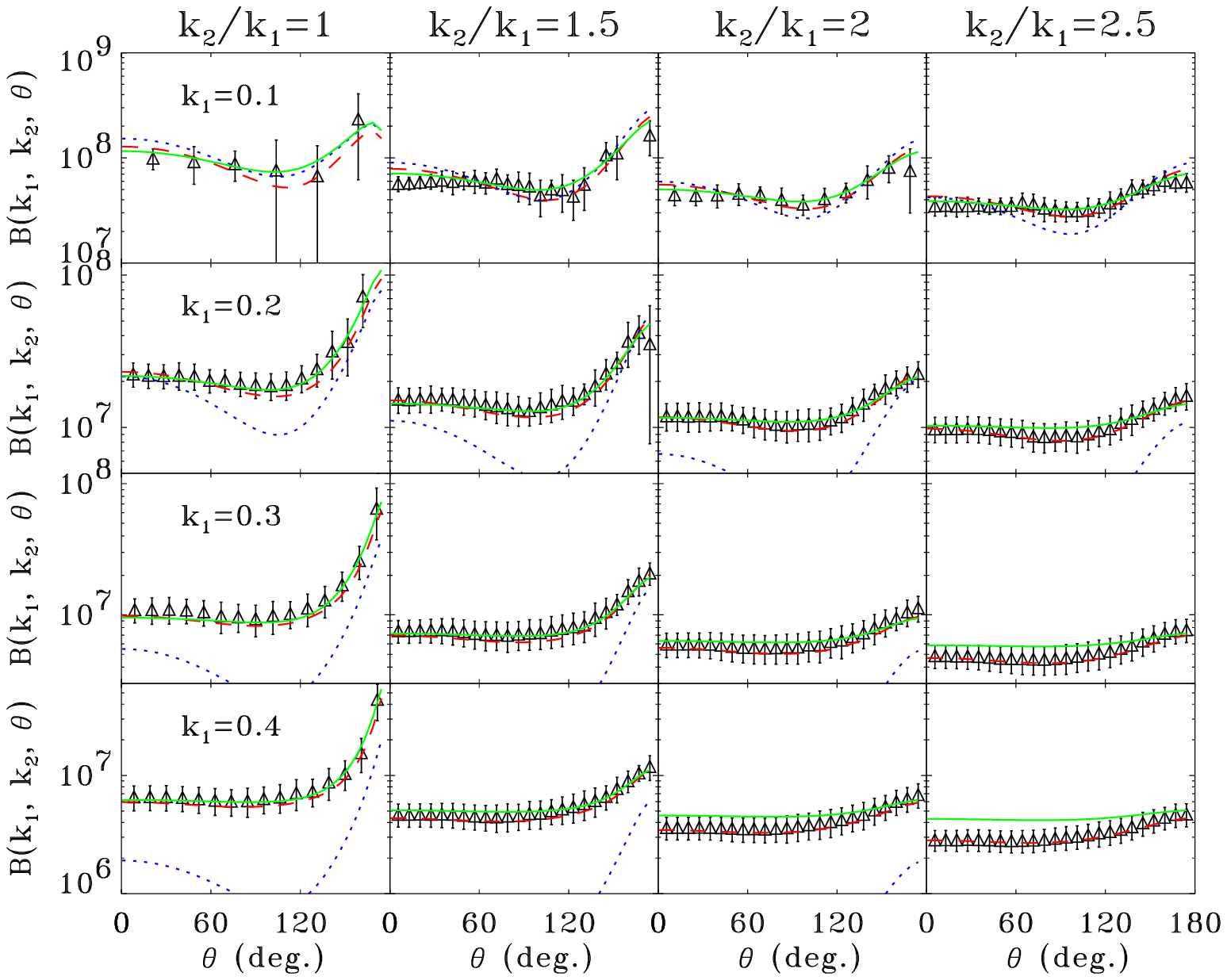}}
\caption{$B(k_1, k_2, \theta)$ at z=0, $\theta=\cos^{-1}\mu_{12}$,
see Fig. (1) for legend and labels.}
\end{figure*}
\begin{figure*}
\resizebox{\hsize}{!}{\includegraphics{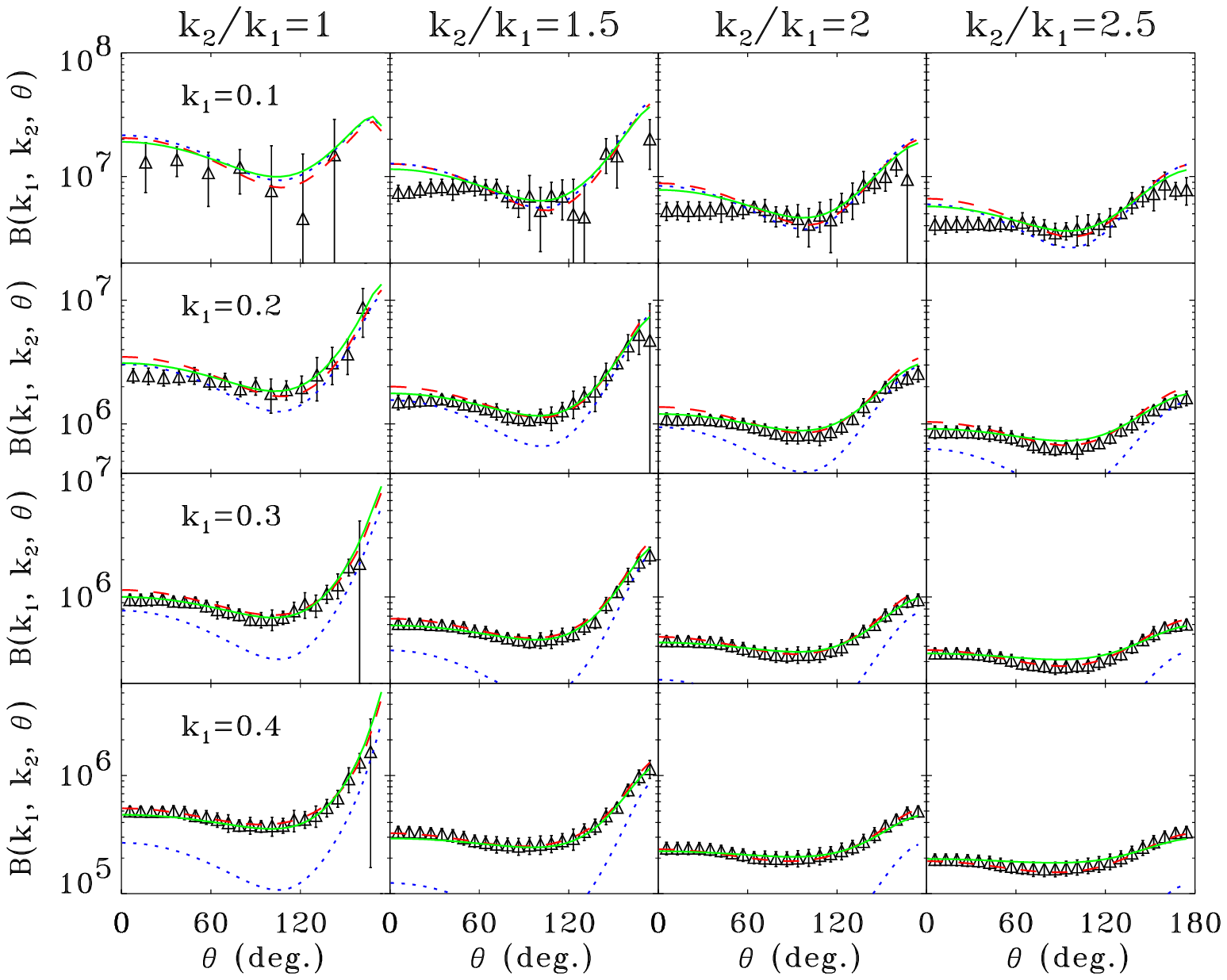}}
\caption{$B(k_1, k_2, \theta)$ at z=1, $\theta=\cos^{-1}\mu_{12}$,
see Fig. (1) for legend and labels.}
\end{figure*}

\subsection{Configuration Re-arrangement}
The three wave vectors defining the bispectrum form a triangular
configuration, and it is well known that there is strong dependence
of bispectrum on the shape of this triangle. If we assume that the
transformation rule defined by Eq.~(\ref{eq:nsr}) holds for
arbitrary $k$, so that a new set of scales $\widetilde{k}$ in
Fourier space is given by
\begin{equation}
\widetilde{k}_{1, 2, 3}=\left[ 1+\Delta_{NL}^2 (k_{1,2,3}) \right]^{-1/3} k_{1, 2, 3} \ ,
\end{equation}
with $\Delta_{NL}^2$ computed by the formula of Smith et al. (2003),
the shape of the original triangle is obviously not preserved. By
the requirement of $\sum {\bf \widetilde{k}}_{1,2,3}=0$, the
transformed triangle has new set of cosines $\widetilde{\mu}$, e.g.
\begin{equation}
\widetilde{\mu}_{12}=-(\widetilde{k}_1^2+\widetilde{k}_2^2-\widetilde{k}_3^2)/(2 \widetilde{k}_1
\widetilde{k}_2) \ .
\end{equation}
Now comes the crucial step in our model. We assume that the
non-linear bispectrum is directly related to the tree-level
bispectrum of the new configuration in Lagrangian space. In the
non-linear regime this relationship is expected to be expressed by a
very complicated functional form, but here we simply conjecture that
we can use the linear expression,
\begin{equation}
B(k_1, k_2, \mu_{12})\propto B_{PT}(\widetilde{k}_1,\widetilde{k}_2, \widetilde{\mu}_{12})\ .
\end{equation}
The remaining ingredient we need is the factor by which the
amplitude of the bispectrum is boosted during the rescaling.

\subsection{Amplitude Boost}
We have tested many schemes to calculate the boost to be applied to
the amplitude, but the best is one involves the idea that a
non-linear fluctuation in Lagrangian space is expressible as a
convolution of the linear density contrast with some filter so that,
in Fourier space,
\begin{equation}
\delta_{NL}(k_L)=w(k_L) \delta_L(k_L)\ ,
\end{equation}
where $w(k)$ is the window function (the Fourier transform of the
filter) and $k_L$ is given by Eq.~(\ref{eq:nsr}). The form of this
convolution is motivated by terms that arise in halo model from
cross-correlation terms in the power spectrum and bispectrum. In
this scenario, the nonlinear growth of the fluctuation in a region
is imagined to consist of two stages: (1) aggregation of clustering
power in Lagrangian space from other regions via convolution; and
(2) translation from the Lagrangian (initial) scale to the Eulerian
(final) scale.

Putting all the elements of Eq. (4) -- (9) together, the expected
nonlinear bispectrum in the recipe is
\begin{equation}
\begin{aligned}
B(k_1, k_2, \mu_{12}) & =w(\widetilde{k}_1)w(\widetilde{k}_2)w(\widetilde{k}_3)B_{PT}(\widetilde{k}_1,\widetilde{k}_2, \widetilde{\mu}_{12}) \\
w(\widetilde{k}_i)&=\left[P_{NL}(\widetilde{k}_i)/P_{L}(\widetilde{k}_i)\right]^{1/2}
\ \ ,\  i=1, 2, 3 \ .
\end{aligned}
\label{eq:model}
\end{equation}
We have to address the issue that, in the model, the possible phase
shifts induced by convolution are not taken into account. This might
have a significant effect in the strongly non-linear regime, since
the bispectrum is a phase-dependent function (Watts \& Coles 2003).

\section{Comparison with simulations}
Four outputs at $z=0, 0.5, 1, 2$ of the VLS (Very Large Simulation)
data provided by the Virgo consortium are used to check our model.
The simulation runs consist of $512^3$ particles in a cubic box of
size $479h^{-1}$ Mpc, and the cosmological parameters adopted are
$\Omega_m=0.3$, $\Omega_v=0.7$, $\Gamma=0.21$, $h=0.7$ and
$\sigma_8=0.9$ (MacFarland et al. 1998). In order to estimate cosmic
variance, the VLS simulation is divided into eight distinct
sub-cubes of half of the original size. The bispectrum is measured
using the method of Scoccimarro et al (1998), and the $1\sigma$
dispersion among the eight measurements is taken as an estimate of
the error bars.

To avoid any bias involved with taking ratios of two statistical
quantities and leakage of the power spectrum into the bispectrum, we
work with the bispectrum itself rather than the reduced bispectrum
$Q_3=B/(P_1 P_2 + P_1 P_3 +P_2 P_3)$ throughout this paper.

\subsection{Equilateral Triangles}
The special case of equilateral triangles is particular
straightforward to interpret because it depends only on one scale.
Moreover, under the effect of the re-arrangement by Eq. (7), an
equilateral triangle does not change shape, so this case should
display self-similar evolution.

In Fig. (1) the bispectrum of equilateral triangles of simulations
is plotted against tree-level perturbation theory, our model Eq.
(10) and the SC2001 for reference. The rescaled tree-level
perturbation theory agrees with simulations remarkably well at
scales $k< 0.7h/$ Mpc at $z=0$. The range of scales in agreement
also keeps increasing at higher redshift, reaching $k\approx 3h/$Mpc
at $z=2$.

In the quasi-nonlinear regime at large scales of $k<0.1h/$Mpc at
$z=0$, Eq. (10) (our model) differs little from the tree-level
perturbation theory as desired, while the suspicious lowering of
SC2001 is clearly seen at low redshifts. Due to the large cosmic
variance, it is not yet possible to tell which model is better.

Our model predicts too much power at scales in the strongly
nonlinear regime, where the bispectrum of the simulations begins to
display a power-law behaviour. The breakdown  can be attributed
either or both of the  two principal weaknesses in our
consideration: either Eq. (8) is simply too rough, or the
correlation between window functions at different values of $k$ is
not negligible.

\subsection{Other Configurations}
The extra complexity of bispectrum over the power spectrum is that
it possesses angular dependence as well as scale dependence.
Comparison of simulations with models are drawn in Fig. (2) and (3)
for $z=0, 1$ respectively.

It is very clear that the our model agrees much better with the
simulations at higher redshift than the alternatives we considered.
It is also observed that Eq. (10) is more accurate when the ratio
$k_2/k_1 \le 2$; within this range the variation with $\theta$ also
matches the simulation results rather well. The major problem of our
model for very tilted triangles is that if we expand $B(k_1, k_2,
\theta)$ with Legendre polynomials then it has a very low quadrupole
moment (Szapudi 2004).

The overall performance of Eq. (10), especially at high redshift, is
comparable with the perturbation theory at one-loop level although
it is far simpler to implement. In fact, as a quick check we realize
that the reduced bispectrum $Q_3$ of our model is very similar to
the one-loop results demonstrated in Scoccimarro et al. (1998), but
emerges in a much more straightforward way using our rescaling
ansatz.

\section{Discussion}
Following the path pioneered by Hamilton et al. (1991) and Peacock
\& Dodds (1996), we deploy a scale transformation argument to
construct a well-behaved model of the bispectrum of dark matter
fluctuations which is valid in the quasi-linear and intermediate
nonlinear regimes. On the basis of the tests we have been able to
perform, the resulting approximation seems to work exceptionally
well, although we are somewhat hampered by numerical limitations. A
full assessment of the precision and reliability of this idea will
have to wait until simulations of even larger size and higher
resolution than the VLS are available.  Nevertheless, the important
point may lie even deeper than the pragmatic usefulness of this as a
simplifying ansatz. The scale transformation of Eq. (1) or (2) may
contain more physical meaning than has been previously thought, and
it may not after all be the case that higher-order correlation
functions in the non-linear regime necessarily pose such fierce
analytical and numerical challenges as has generally been asssumed.

The tree-level bispectrum is fully determined by the linear power
spectrum, while the success of our model seems to tell us that the
nonlinear bispectrum is governed by the nonlinear power spectrum. It
therefore seems possible that having the power spectrum at hand, one
can accurately derive the bispectrum of a broad range of
configurations using only the knowledge that it was generated by
gravitational physics.

It is well known that a Gaussian random field is fully described by
its power spectrum, but does a non-Gaussian field evolved by gravity
from Gaussian initial conditions also possess this feature at some
level? It will be very interesting to test this idea by seeing if
the trispectrum can also be obtained by rescaling  tree-level
perturbation theory. More fundamentally, if gravitational
interactions do manage to act in such a way then why is it that the
simple scale transformation, Eq. (2), manages to capture so much
complicated physics?

\section*{Acknowledgment}
JP acknowledges the fellowship of the One Hundred Talents program of
CAS, this work is also partly support by NSFC under grant 10643002
and by PPARC through grant PPA/G/S/2000/00057. IS acknowledges
support from grants NSF AST02-06423, NSF AMS04-0434413 and NASA
NNG06GE71G. The simulations in this paper were carried out by the
Virgo Supercomputing Consortium using computers based at Computing
Centre of the Max-Planck Society in Garching and at the Edinburgh
Parallel Computing Centre.

\end{document}